\documentclass[submission,copyright,creativecommons]{eptcs}
\providecommand{\event}{FMAS 2022}

\usepackage{iftex}
\usepackage{wrapfig}
\usepackage{graphicx}

\usepackage{amsmath}
\usepackage{amsfonts}
\newcommand{\relationTwo}{R'}
\newcommand{\relation}{R}

\ifpdf
  \usepackage{underscore}         
  \usepackage[T1]{fontenc}        
\else
  \usepackage{breakurl}           
\fi

\newcommand{\acronym}{KANOA}

\title{Scheduling of Missions with Constrained Tasks \\for Heterogeneous Robot Systems}
\author{Gricel V\'azquez \qquad\qquad Radu Calinescu 
\institute{Department of Computer Science\\University of York\\
York, UK}
\email{gnvf500@york.ac.uk \qquad radu.calinescu@york.ac.uk}
\and
Javier C\'amara
\institute{Department of Computer Science\\University of M\'alaga\\
M\'alaga, Spain}
\email{javier.camaramoreno@york.ac.uk}
}
\def\titlerunning{Scheduling of Missions with Constrained Tasks for Heterogeneous Robot Systems}
\def\authorrunning{G. V\'azquez, R. Calinescu \& J. C\'amara}

\begin{document}
\maketitle

\begin{abstract}
We present a formal tas\underline{k} \underline{a}llocatio\underline{n} and scheduling appr\underline{oa}ch for multi-robot missions (\acronym). \acronym\ supports two important types of task constraints: task ordering, which requires the execution of several tasks in a specified order; and joint tasks, which indicates tasks that must be performed by more than one robot. To mitigate the complexity of robotic mission planning, \acronym\ handles the allocation of the mission tasks to robots, and the scheduling of the allocated tasks separately. To that end, the task allocation problem is formalised in first-order logic and resolved using the Alloy model analyzer, and the task scheduling problem is encoded as a Markov decision process and resolved using the PRISM probabilistic model checker. We illustrate the application of \acronym\ through a case study in which a heterogeneous robotic team is assigned a hospital maintenance mission.

\end{abstract}

\section{Introduction}
Multi-robot systems (MRS) are increasingly used in domains ranging from medical and emergency assistance~\cite{askarpour2021robomax} to  inspection of critical infrastructures~\cite{sukkar2019multi}.
Nevertheless, the specification of complex missions with constrained tasks, the allocation of these tasks to the robots, and the scheduling of allocated tasks for each robot continues to pose significant challenges to the use of MRS in these applications~\cite{fox2003pddl2,tran2017robots,wang2020learning}. 

Our paper introduces \acronym, a new MRS tas\textbf{k} \textbf{a}llocatio\textbf{n} and scheduling appr\textbf{oa}ch that uses a combination of formal methods to solve what Tran et al.~\cite{tran2017robots} describe as ``\textit{the joint problem of deciding what tasks to perform, when, and with what  resources, to achieve a set of goals}''.
Task allocation and scheduling can be tackled in two ways: as a single, monolithic problem~\cite{gavran2017antlab}, or separately, as two interrelated sub-problems~\cite{gombolay2013fast}. \acronym\ does the latter. As in~\cite{gombolay2013fast}, we posit that this enables a separation of concerns that not only increases scalability, but also simplifies the enforcement of a broad range of mission constraints and the achievement of optimization objectives.

Task allocation and scheduling is among the problems that can be approached using formal methods~\cite{woodcock2009formal}, which offer underlying logics for the unambiguous specification of missions, and behavioural models that are able to capture important MRS mission characteristics such as task interdependence~\cite{garcia2020promise}. 
However, the adoption of formal methods in real MRS scenarios has been limited due to the complexity of formalising MRS missions and their requirements into the required modelling paradigms and logic formulae, respectively~\cite{yu2021distributed,carreno2020task}. To address this limitation, \acronym\ (i)~supports the specification of MRS missions with their tasks, task constraints and requirements in a domain-specific language, and (ii)~automates both the allocation of tasks to robots (by using the model analyzer Alloy~\cite{alloy}) and the scheduling of the allocated tasks (by using the probabilistic model checker PRISM~\cite{kwiatkowska2011prism}). PRISM schedules the tasks allocated to each robot by solving the ordering of the tasks through synthesising a Markov decision process policy that minimises the time needed to complete the tasks. 

The rest of the paper is organised as follows. After providing a motivating example in Section 2, we present our KANOA approach in Section 3, and its implementation and evaluation
in Sections 4 and 5, respectively. Section 6 discusses related work, and Section 7 a short summary.

\section{Motivating Scenario} 
\label{sec:motivating}

\begin{wrapfigure}[18]{r}{0.5\textwidth}
    \vspace{-45pt}
    \centering
    \includegraphics[width=1\linewidth,]{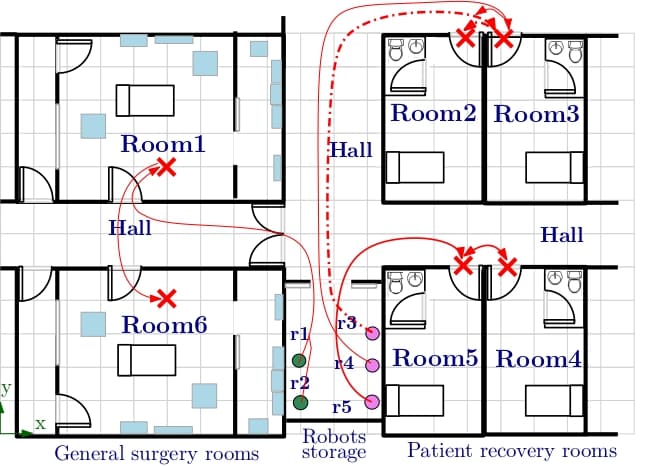}
    \caption{Floor plan of hospital area with six rooms serviced by robots {\sf r1-r5}. Crosses mark the locations that robots must reach to start a task in each room, and blue rectangles denote medical equipment (inspired by \cite{nollert2008planning}). The red lines shows a possible robots' schedule to clean rooms 2 to 4, and to move equipment in rooms 1 and 6.}
    \label{fig:workspace}
\end{wrapfigure}

We will illustrate our approach in the context of a hospital environment, in which patient recovery rooms must be kept clean, and medical equipment has to be moved around to perform daily operations (Fig.~1). We introduce two types of tasks: atomic tasks, {\sf AT}; and compound tasks, {\sf CT}.
The MRS mission consists of six tasks: 
moving medical equipment in Room1 (task \textsf{M1}) and Room6 (task \textsf{M2}); and cleaning patient rooms 2--5 (tasks \textsf{M3} to \textsf{M6}). 
Moving medical equipment is a type of task ({\sf AT1}) that must be performed by two robots. 
Cleaning a patient's room~({\sf CT2}) requires notifying the patient~({\sf AT4}) and cleaning the room~({\sf CT1}), in that specific order. 
Finally, cleaning a room comprises floor cleaning~({\sf AT2}) and sanitising~({\sf AT3}). 
Five robots are available: two cleaner robots~({\sf r1} and \textsf{r2}) and three pick-and-place robots~({\sf r3-r5}).

\begin{wrapfigure}[17]{r}{0.53\textwidth}
    \centering
    \vspace{-10pt} 
    \includegraphics[width=1\linewidth]{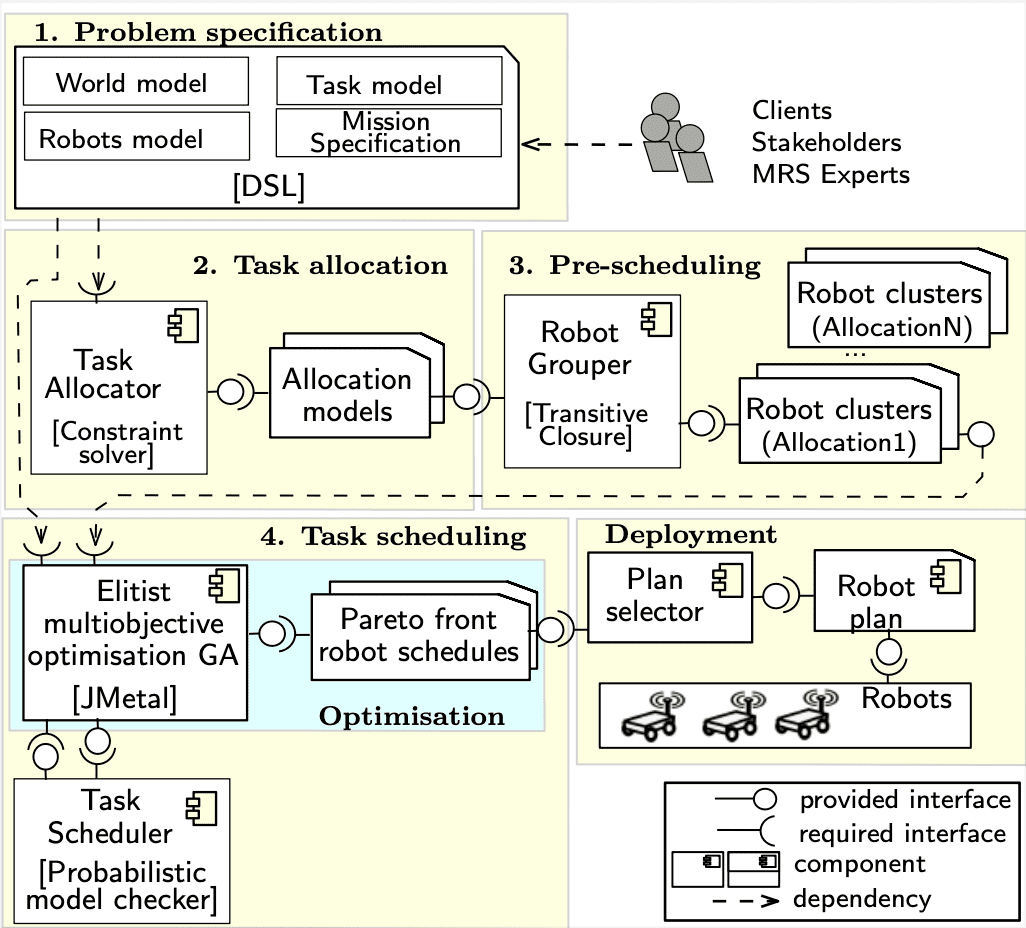}
    \caption{The four-stage \acronym\ approach}
    \label{fig:approach}
\end{wrapfigure} 

In addition to the constraints imposed by the ordering of tasks and the tasks requiring more than one robot, the mission must be completed within 100~time units, and the robots that are used to perform the mission cannot spend more than 30~time units idle. Furthermore, the tasks must be allocated and scheduled such that the total robot idle time and travelling time are minimised, and 
the probability of succeeding with the mission is maximised. An example of a synthesized plan (discussed later in the paper) is shown by the red arrows from Fig.~\ref{fig:workspace}.

\section{\acronym\ Approach}
\label{sec:approach}

\acronym\ employs the four-stage process from  Fig.~\ref{fig:approach}. In stage~1, the \acronym\ domain-specific language (DSL) is used to specify the 
world model, robots model, task model and mission that together form the problem that \acronym\ needs to solve. 
In stage~2, \acronym\ solves the allocation of tasks to robots, producing multiple allocation models, all of which comply with the robot capabilities. Stage~3, pre-scheduling, groups robots that share constrained tasks, so that their tasks are scheduled together. Finally, the task scheduling stage generates a set of feasible MRS plans that satisfy all task constraints and are Pareto optimal with respect to the optimisation objectives from the mission specification. The user can select one of the synthesized plans to be deployed to the physical robots, as shown in the deployment stage. \acronym\ focuses on the four \acronym\ stages as detailed below. 

\subsection{Stage~1: Problem specification}
\label{sec:approachSpecificationDSL}

\begin{figure}
\centering
\includegraphics[width=0.89\linewidth]
{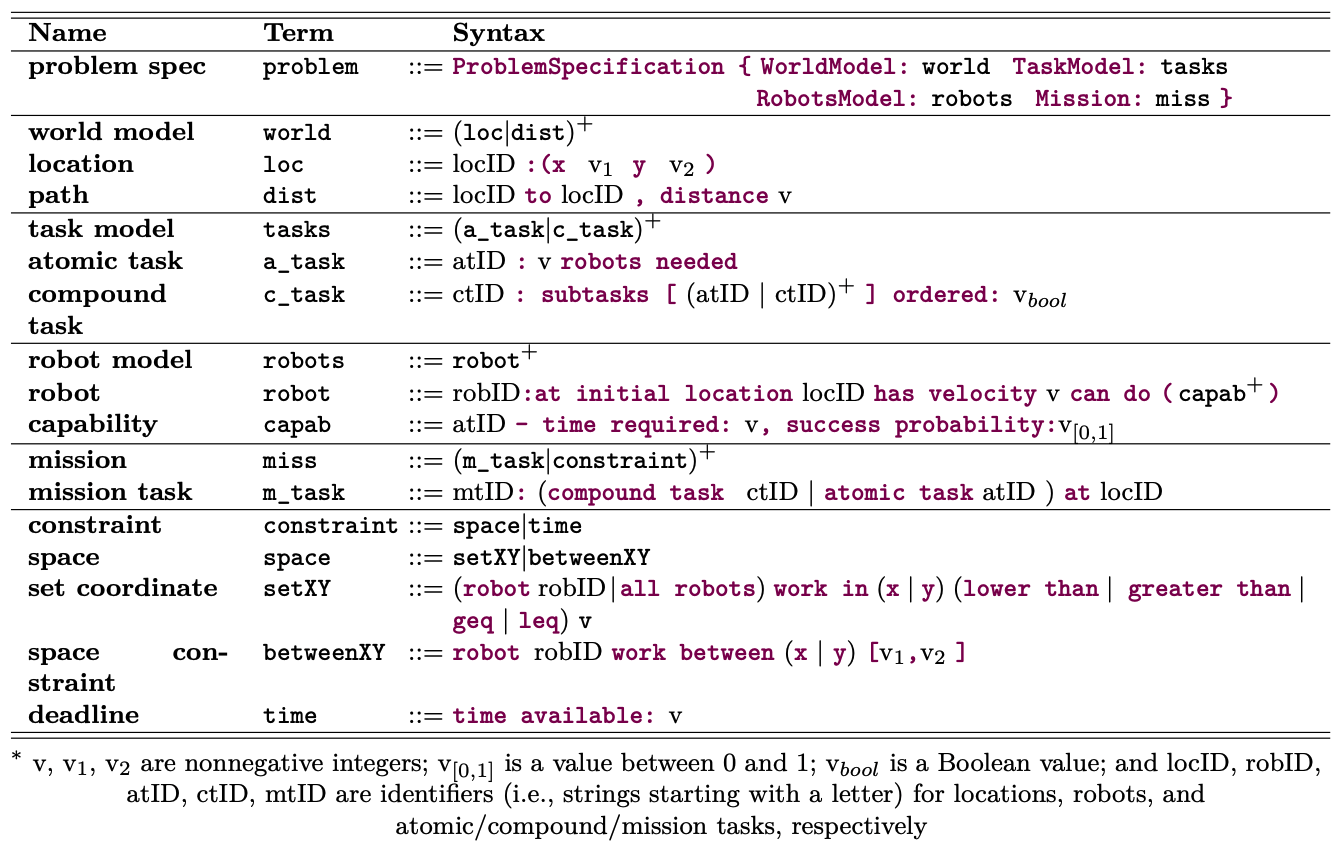}
\caption{\acronym\ problem specification DSL. Based on the EBNF metasyntax notation~\cite{standard1996ebnf}.}
\label{figure:dslsyntax}
\end{figure}

We use the user-friendly DSL with the syntax from Fig.~\ref{figure:dslsyntax} for the unambiguous specification of the \acronym\ task allocation and scheduling problem. In line with established conventions, alternative items are divided by $|$ and may be enclosed in brackets (). Items that can appear one or more times are followed by a plus~+.

A \acronym\ \textbf{problem} specification comprises a \textbf{world} model, a \textbf{task} model, a \textbf{robot} model, and a \textbf{mission} specification: 
\begin{itemize}
    \item The \textbf{world} model consists of a sequence of locations \textbf{loc} and path (distances) \textbf{dist}, where each location is defined by an identifier \textbf{locID} and its \textbf{x} and \textbf{y} coordinates, and each path \textbf{dist} specifies the all of the distance between two locations.
    \item The \textbf{tasks} model defines a set of \emph{atomic tasks} and \emph{compound tasks}. An atomic task (\textbf{a\_task}) needs to be performed in one go, by a predefined number of \textbf{robots needed}. In contrast, a compound task (\textbf{c\_task}) comprises a set of \textbf{subtasks}, each of which is an atomic or a compound task. The \textbf{subtasks} of a compound task may need to be performed in the specified order.
    \item The \textbf{robots} model specifies, for each robot available, its \textbf{initial position}, its (mean) \textbf{velocity} and its list of \emph{capabilities}, i.e., the identifiers \textbf{atID} of the atomic tasks that the robot \textbf{can do}, and the \textbf{required time} and \textbf{success probability} associated with the task execution by the robot\footnote{The success probabilities differ from robot to robot based on age, manufacturing differences, etc.; and calculated in advance after N runs of the physical or simulated robots}.
    \item The mission specification (\textbf{miss}) comprises a sequence of mission tasks, \textbf{constraint}s. 
    A mission task (\textbf{m$\_$task}) specifies a \textbf{compound task} or \textbf{atomic task} (from the \textbf{task} model), which needs to be performed \textbf{at} a location specified by \textbf{locID}. \textbf{Constraint}s define boundaries within which a specific \textbf{robot} \textbf{robID} or  \textbf{all robots} must reside at all times, or the \textbf{time available} to complete the mission. 
\end{itemize}

\begin{figure}[t]
\centering
    \includegraphics[width=0.85\linewidth]{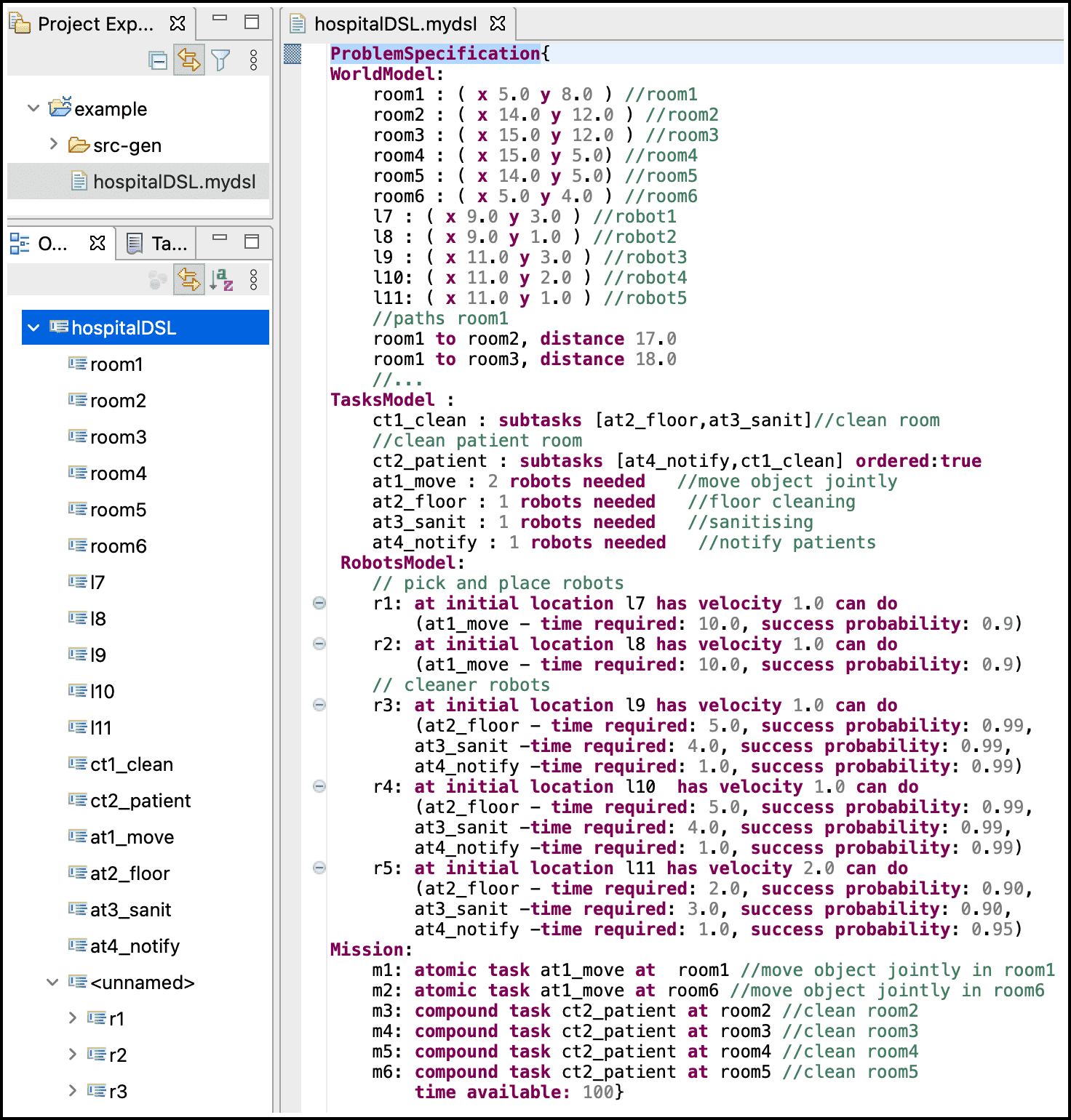}
    \caption{\acronym\ tool screenshot showing DSL encoding of the hospital MRS mission from Section~\ref{sec:motivating}}
    \label{fig:toolDSL}
\end{figure}

\medskip 
\noindent
\textbf{Example 3.1.}
\label{sec:exampleHospitalDS}
Figure \ref{fig:toolDSL} shows the \acronym\ problem specification of the hospital scenario from our motivating example. As detailed in  Section~\ref{sec:motivating}, this scenario comprises: (a)~six rooms with their (entry point) coordinates, initial locations for the five robots, and all distances between these; (b)~two compound tasks and four atomic tasks; (c)~two pick-and-place robots (r1 and r2) and three cleaner robots (r3--r5); (d)~the mission that the robots are required to carry out (within 100 time units), consisting of moving medical equipment in rooms 1 and 6, and cleaning the patient rooms 2--5. Pick-and-place robots r1 and r2 require 10 time units to reorganise the medical equipment in a surgical room. Cleaner robots r3--r5 have a higher probability of success compared to r5, but r5 travels twice as fast as r3 and r4.

\subsection{Stage 2: Task allocation}
\label{sec:approach-allocation}

The multi-robot task allocation problem involves the optimal partitioning of the tasks of a mission among the available robots. This problem is very challenging due to the heterogeneity of robots, the unreliability of sensors and actuators, temporal and spatial task constraints, the complexity of task dependencies, and the search for an optimal allocation~\cite{khamis2015multi}. The \acronym\ Task Allocator (see Fig.~\ref{fig:approach}) performs the allocation of tasks to robots by using the Alloy Analyzer constraint solver~\cite{alloy} to  generate feasible allocations through reasoning about the capabilities of each robot, and the spatial constraints from the mission specification.

\subsubsection{Background}
\label{sec:alloy}

Alloy is a formal specification language (supported by the {\em Alloy analyzer} tool) that has its roots in the Z specification language~\cite{alloy}. It supports the formal description of the structural properties of systems. Alloy is defined in terms of simple relational semantics, using constraint solver notation, and employing constructs that are common in object-oriented notations. We can indicate the existence of a set (of atoms) for robots, which in Alloy is specified using {\em signatures}:

\medskip
{\small
\hspace*{1cm}{\sf \bf abstract sig} Robot \{hascapability: {\sf  \bf some} Capability\}
}

\medskip\noindent
In this signature, {\sf Robot.hascapability} denotes a relation with a non-empty set of capabilities that a robot possesses. The keyword `{\sf abstract}' indicates that new signatures can {\em extend} this signature, inheriting all of its characteristics. 
In our example, we can, for instance, create concrete robots and capabilities (e.g., {\sf r1}, {\sf r2} and {\sf c1}, {\sf c2}) that instantiate their corresponding abstract signatures:

\medskip
{\small
\hspace*{1cm}{\sf \bf sig} r1, r2 {\sf\bf extends} Robot \{ ... \}
}

\medskip
Alloy models can incorporate relational logic sentences in the form of predicates and {\em facts} (i.e., expressions that always have to be satisfied). These constructs place explicit constraints on the model. 
Hence, when the Alloy analyzer searches for instantiations that satisfy the structural properties described in the Alloy model, it discards any which violate any fact. 
In our example, we can write a fact which makes sure that, if a capability {\sf c} belongs to a specific robot {\sf r}, that robot indeed has {\sf c} among its capabilities (and the other way around):

\medskip
{\small
\hspace*{1cm} {\sf {\bf fact} \{ {\bf all} c:Capability, r:Robot $|$ r in c.belongsto {\bf $<$=$>$} c in r.hascapability \} } 
}

\medskip
\noindent
The Alloy analyzer can look for examples of structures that satisfy all of the relational constraints in the model within a finite scope that explicitly sets the maximum number of atoms of each signature to be considered by a solution.

\subsubsection{Task allocation problem}
\label{sec:Task allocation problem}
\begin{figure*}[bt!]
\centering
\includegraphics[width=0.9\linewidth,height=0.27\linewidth]
{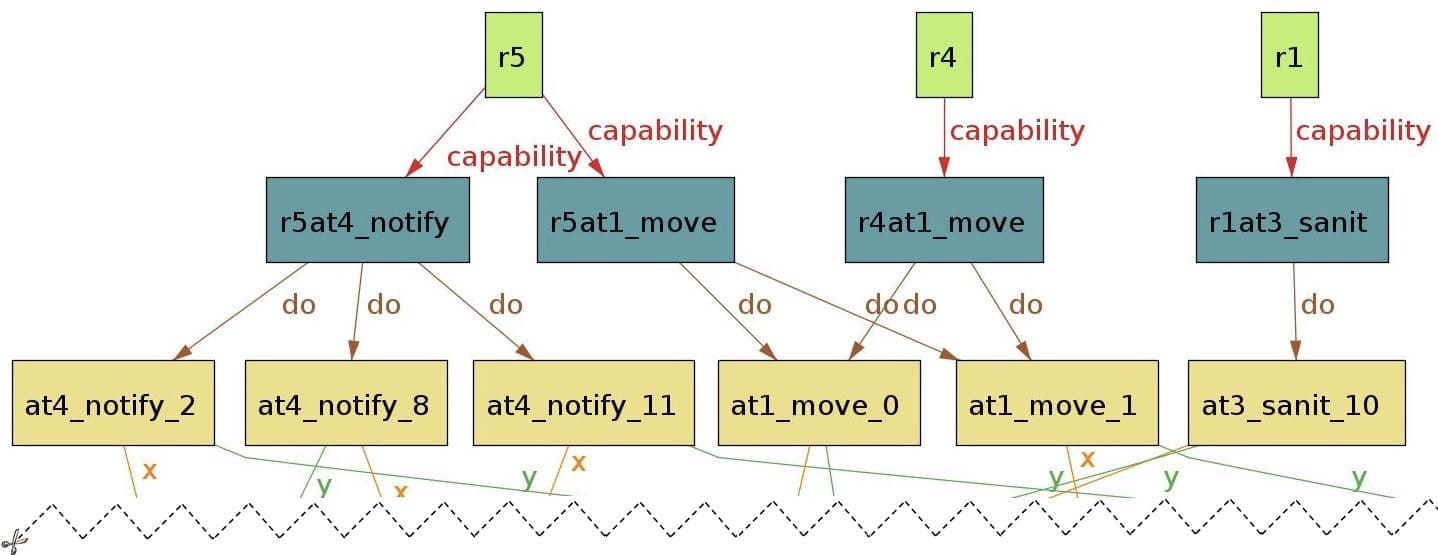}
\caption{Allocation generated by Alloy Analyzer.}
\label{fig:allocationAlloy}
\vspace{-5mm}
\end{figure*}

We define the task allocation problem in the Alloy declarative language with abstract signatures for atomic tasks, robot capabilities and robots. There are four facts associated with these signatures. First, a robot has its own set of capabilities, meaning that each capability belongs to only one robot. Second, if a robot appears in the allocation~(see Figure~\ref{fig:allocationAlloy}) it must have assigned tasks, so that robots that are not used are ignored. Third, only capabilities that have allocated tasks are shown. Finally, each robot appears once or, if not selected by the allocator, does not appear at all.

For the declaration of atomic tasks, we model a second abstraction extending the $AtomicTask$ abstract signature. These are atomic tasks specified in the problem specification. The instances of these tasks are atomic tasks required by the mission (recursively fetched from the \textit{tree-like} structure of the tasks model). We explicitly define the maximum number of robots and capabilities, the number of atomic tasks and each atomic task, to generate $N>0$ possible allocations, where $N$ is a \acronym\ configuration parameter.

\begin{wrapfigure}[15]{l}{0.5\textwidth}
    \centering
    \vspace{-10pt}
\includegraphics[width=\linewidth]{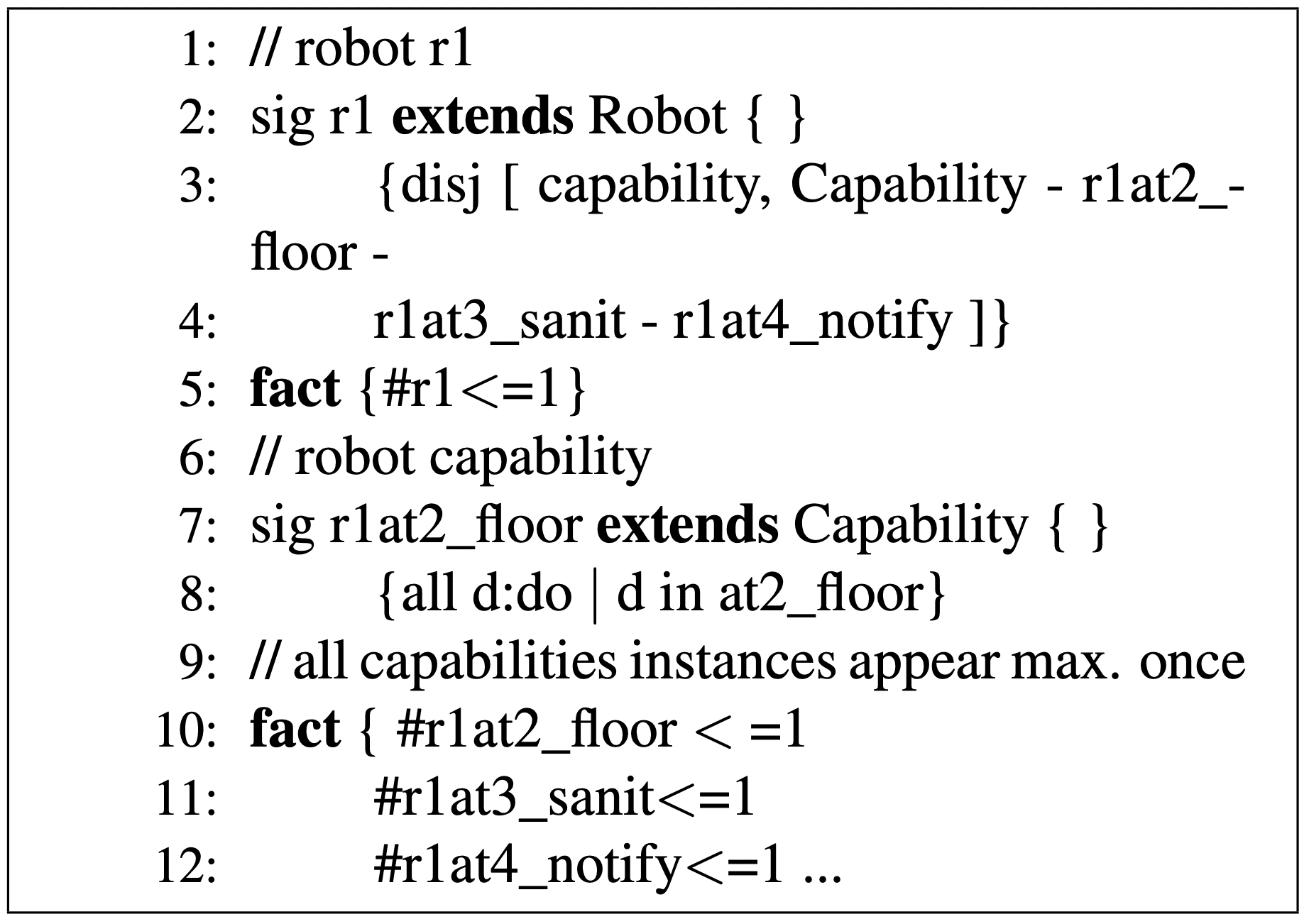}
\vspace{-13pt}
    \caption{Path from the initial location of robot r2 to room 6 found using PRM}
    \label{fig:alloy3}
\end{wrapfigure}

Figure \ref{fig:allocationAlloy} shows part of an allocation found by Alloy Analyser. At the top are three of the robots, $r5, r4$ and $r1$. Only capabilities needed from the robots for the allocated tasks are shown, for instance, $r1$ only shows one capability $r1at3\_sanit$. Notice that robots $r4$ and $r5$ have the same allocated task $at1\_move\_0$ as it is a joint task. Links (x,y) at the bottom of the figure point to the locations where each atomic task must be done.

\medskip
\noindent
\textbf{Example 3.2}
An example of a robot instance based on our motivating example is shown in Figure \ref{fig:alloy3}, lines 1-4. In this case, robot $r1$ has capabilities $r1at2\_floor, r1at3\_sanit$ and $r1at4\_notify$, modelled as the disjoint set of these tasks robot capabilities and the set of all capabilities. The number of robots required is written as a fact (line 5).
A robot capability instance has the constraint of only performing atomic tasks in a specific scope. For example, robot $r1$ with capability $r1at2\_floor$ can only perform $at2\_floor$ type of tasks, as specified in Figure \ref{fig:alloy3} lines 6-8. We add a constraint to show a capability instance once, or 0 if it has no allocated tasks (lines 9-13).
 Due to space constraints, we refer the user our GitHub repository for the complete Alloy code for the motivating example~\cite{githubRepository}.

\subsection{Stage~3: Pre-scheduling}

For each allocation found by \acronym, we reduce (where possible) the complexity of the scheduling models, by creating separate models for each \textit{group} of robots that share task dependencies and do not share any robot with the other groups. We call this stage \textit{pre-scheduling}, as shown in Figure~\ref{fig:approach}.

\begin{figure}[t]
\centering
\includegraphics[width=1\linewidth]{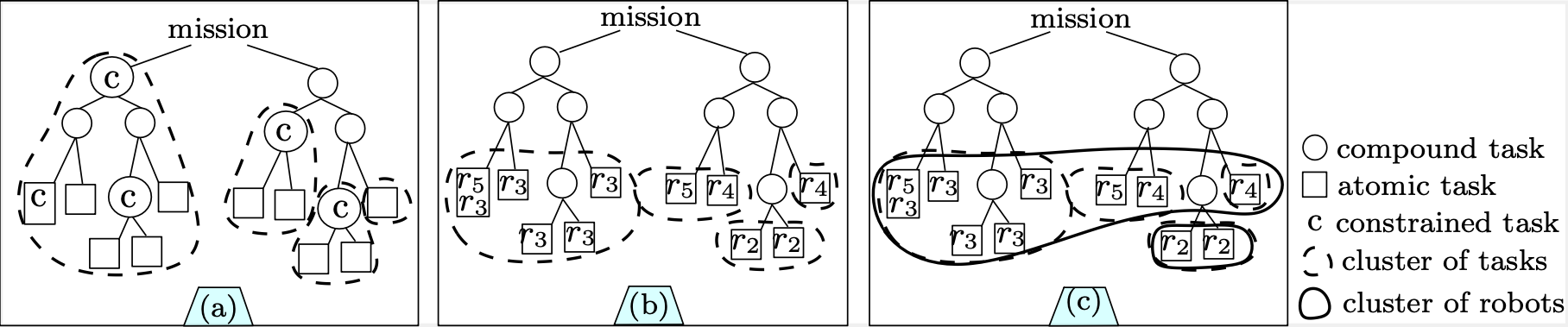}
\caption{Transitive closure of robot interdependence relation: (a)~atomic tasks clustered by breath-first until a constrained task is found; (b)~for each subtree, the robots in the current allocation are obtained; (c)~the transitive closure returns clusters of robots---in this example, there are two robot clusters, i.e., $\{r3,r4,r5\}$ and $\{r2\}$.}
\label{fig:transitiveclosure}
\end{figure}

Let $T$ be the tree of tasks from the mission specification to the atomic tasks. First, \acronym\ computes the Breadth-first search $BFS$ algorithm pruning the tree every time that a constraint (joint or consecutive task) is found, or a leaf node (atomic task) is reached, returning multiple subtrees, as shown in Fig.~\ref{fig:transitiveclosure}a.

\begin{equation}
    BFS(T) = \{subtree_1, subtree_2, ...\}
\end{equation}

Let $A_i$ be the $i^{th}$ allocation of tasks to robots found by Alloy, and $Rob_i$ the set of all robots in $A_i$; for example, the task allocation in Figure~\ref{fig:transitiveclosure}.b has robots $Rob_i=[r2,r3,r4,r5]$. For each $subtree_j\in BFS(T)$, we define the set of robots associated with its leaf nodes by,
\begin{equation}
    Rob_{ij}(A_i, subtree_j) \subseteq Rob_i
\end{equation}

\noindent
as shown in Fig.~\ref{fig:transitiveclosure}b. We define a reflexive binary relation $R$ between robot $r_a$ and $r_b$ as,
\[
     r_a\,R\,r_b= 
\begin{cases}
    1,  &   \text{if } \exists subtree\in BFS(T) \cdot
    r_a, r_b \in subtree\\
    0,  &   \text{otherwise}
\end{cases}
\]

Finally, we define the \textit{transitive closure} of $R$ as a reflexive relation $\relationTwo$ such that $r_a\relationTwo r_k$ if there is a sequence $(r_a,r_b,...,r_j,r_k)$ of the members of the domain of $\relation$ such that $r_a\relation r_b,\ldots,r_j\relation r_k$, as shown in Fig.~\ref{fig:transitiveclosure}c.
The relation $R$ may be represented as a Boolean matrix $M$ (where $M(r_a,r_b)=r_aRr_b$), and $\relationTwo$ as matrix $M'$ computed from $M$ using $Warshall$'s algorithm or variations~\cite{warren1975modification}.

\subsection{Stage 4: Task scheduling}
\label{sec:approach-scheduling}

\acronym\ uses probabilistic model checking for the scheduling of individual robot plans. In short, it models the scheduling problem as a Markov decision process (MDP), finding a policy that optimises a set of objectives under a set of constraints. A policy is a solution to the non-determinism actions, assessing if a robot should do the next task or wait. Hence, finding a policy means finding a feasible schedule.

\subsubsection{Background}

\textbf{Markov decision processes (MDPs)} support the modelling of probabilistic and nondeterministic behaviour~\cite{forejt2011automated}. An MDP is defined as a tuple $\mathcal{M}=(S,s_i,A,\delta_\mathcal{M},L)$, where $S$ is the finite set of states and $s_i\in S$ an initial state; $A$ is the set of actions and  $\delta_\mathcal{M} : S\times A \rightarrow Dist(s)$ is a partial probabilistic function returning a distribution over $S$ when an action $a \in A$ is taken from $s\in S$; and, $L: S\rightarrow 2^{AP}$ is a labelling function assigning a set of atomic prepositions from a set $AP$ to each state $s\in S$. 
\textit{Reward} structures that associate a quantity to each action choice can be used, for example, to track the energy consumption, or  the total  travelling distance of a robot.

\medskip\noindent
An MDP \textbf{policy} resolves the non-deterministic choice of an action in the states of an MDP~\cite{forejt2011automated}. \acronym\ uses $deterministic-memoryless$ policies, which involves the selection of a fixed action whenever the same state is reached~\cite{gerasimou2021evolutionary,forejt2011automated}.

\medskip\noindent
\textbf{Probabilistic computational tree logic (PCTL)} augmented with rewards is used to define qualitative and quantitative formulae related to MDP rewards and probabilities that can be assessed via probabilistic model checking. A \textit{state PCTL formula} $\Phi$ and a \textit{path PCTL} formula $\Psi$ are defined over a set of atomic prepositions $AP$ by the grammar:
\begin{equation}
    \begin{array}{l}
    \Phi := true|\alpha | \lnot \Phi | \Phi \land \Phi | \mathcal{P}_{\bowtie p}[\Psi] | 
    \mathcal{R}_{\bowtie r} [F \Phi] \\ 
    \Psi := X\Phi | \Phi U \Phi | \Phi U^{\leq k} \Phi
  \end{array}
\end{equation}
\noindent
with atomic preposition $\alpha\in AP$, probability bound $p\in [0,1]$, reward bound $r\in \mathbb{R}_0^+$, time step bound $k\in \mathbb{N}_{>0}$, and $\bowtie\ \in\{\geq,>,<,\leq\}$.
PCTL \textit{semantics} is defined by the satisfaction relation $\models$. For a state $s\in S$ of an MDP $\mathcal{M}$, we describe a relation $\mathcal{M},s \models \Phi$ as ``\textit{the formula $\Phi$ holds on $s$}''. Hence, the following relations are valid: always $s\models true$; $s\in \neg\Phi$ iff $\neg(s\models \Phi)$; $s\models \alpha$ iff $\alpha\in L(s)$; and $s\models\Phi_1\land\Phi_2$ iff $s\models\Phi_1$ and $s\models\Phi_2$. The \textit{until} formula $\Phi_1U\Phi_2$ holds for a path iff $\Phi_1$ holds in the first $i$ states and $\Phi_2$ holds in the state $i+1$. The $next$ formula $X\Phi$ holds if $\Phi$ is satisfied in the following state. Formulae with the probability symbol $\mathcal{P}$, have their semantics defined over all policies $\sigma$ of $\mathcal{M}$, such that $\mathcal{P}_{\bowtie p}[\Psi]$ is the probability that paths starting at a defined initial state satisfy a path property $\Psi$ with probability $\bowtie p$ for all policies. Replacing $\bowtie p$ with $min=?$ or $max=?$ specifies the calculation of the minimum or maximum probability over all the MDP policies. Similarly, for reward formulae, $\mathcal{R}_{\bowtie r}[F\Phi]$ (where $F$ represents a reachable state in the \textit{future}) holds if the expected reward accumulated before reaching a state satisfying $\Phi$ is $\bowtie p$ for all policies. Replacing `$\bowtie r$' with `$min=?$' or `$max=?$' specifies the calculation of the minimum or maximum reward over all the MDP policies. We refer the reader to~\cite{forejt2011automated,bianco1995model,hansson1994logic} for further details.

\medskip\noindent
\textbf{PRISM modelling language}
\acronym\ MDPs are defined in the high-level modelling language of the probabilistic model checker PRISM~\cite{kwiatkowska2011prism}. A model specified in this language comprises several interacting reactive modules. Each such module consists of a set of finite-valued state variables, and a number of transitions (i.e., commands) modifying these variables. A \textit{transition} has the generic form\\[2mm]
\hspace*{2cm}\text{[$<$action$>$]} $<$guard$>$ $\rightarrow$ $<$prob$>$:$<$update$>$ +...+ $<$prob$>$:$<$update$>;$\\[2mm]
and comprises an
optional \textit{action} label, a guard and state updates with transition probabilities assigned. An \textit{action} is a label that allows synchronisation between modules. A \textit{guard} is a predicate over the variables of all modules. An \textit{update} modifies the module variables if the transition is taken ($\rightarrow$) with a probability of \textit{prob}. In an MDP, non-deterministic actions are taken when two transitions in the same or different modules have overlapping guards.

\subsubsection{Multi-robot task scheduling problem}
\label{section: Multi-robot task scheduling problem}

\begin{figure*}[t!]
    \centering
    \includegraphics[width=0.9\linewidth]{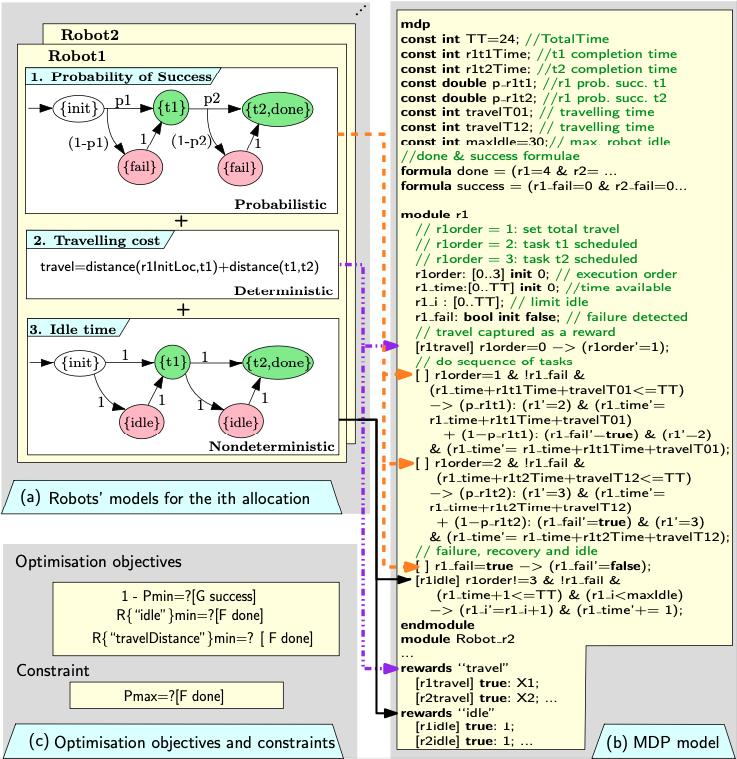}
    \caption{Model assembly for scheduling the tasks of interdependent robots under task constraints and multiple optimisation objectives: (a)~representation of the models---(1)~maximisation of probability of success, (2)~minimisation of the travelling distance, and (3)~minimisation of the idle time; (b)~MDP assembly with the actions and state variables from (a); and (c)~the three optimisation objectives. The bottom left shows the definitions for A and B for the MDP in (b).}
    \label{fig:compositionMDP}
\end{figure*}

We use MDPs encoded in the PRISM language to capture the \acronym\ multi-robot task scheduling problem.
Informally, we define this problem as finding a sequence of possible \textit{robot actions}~(execute a task, travel, or stay idle) such that, for each robot, all of its allocated tasks from Section~\ref{sec:Task allocation problem} are completed with an optimal level~(minimising the completion time, travelling cost and probability of success), preserving any required task ordering and joint execution of tasks across the schedules of all robots.

To this end, we first generate feasible permutations of tasks that satisfy the tasks constraints. This is done by randomly selecting a task that is not part of any constrained task, or the first task to be done in a task tree of ordered tasks.
We use MDPs to model the robots' behaviour~\cite{kwiatkowska2011prism}. For an allocation $a$ with robot $r_i$, a permutation of its tasks is given by $p = a.modules[r_i]$. Each permutation is modelled as a PRISM module with state variables $s_{r_i}$ defined by the tuple,

\begin{equation}
    s_{r_i} = (r_iorder, r_itime, r_iidle,  r_ifail,r_itasks)
\end{equation}
where $r_iorder$ tracks the order in which tasks are executed and the travelling action~(further explained later); $r_itime$ tracks the current time updating every time a robot travels to another location or a task is completed and $r_idle$ the time that the robot has spent idling; $r_itravel$ contains the information of the robots travelling time; $r_ifail$ is a boolean flag triggered when the robot fails performing a task; and $r_itasks$ is a set of variables associated to task constraints~(see Section~3.4.3). An MDP state is given by the composition of each robot module variables $s=(s_{r1},s_{r2},...,s_{rn})$.

The MDP modules can be explained as the composition of three sub-models, as shown in Figure \ref{fig:compositionMDP}a. In the first model, a probabilistic choice is made between succeeding with a task or failing with a small probability. From the robot's initial position, where atomic proposition $init$ is true, a robot travels and completes its first task with probability $p1$ (state with atomic preposition $t1$), and with probability $1-p1$ fails. If a robot fails, the system recovers and continues with the next task\footnote{We consider the probability of recovery of $1$. However, it can be modelled as multiple recovery modes, for instance, with a probability $p1$ the robot can retry and succeed, with probability $p2$ it asks for help from a human, and with probability $p3$, the robot sends an alert that the task failed and needs to be rescheduled.} until all tasks are completed (label $done$).

Second, as shown in Section~2 from Figure \ref{fig:compositionMDP}a, the travelling cost is computed as the sum of the distances from the initial location of the robot to the location of its last task, passing through the locations of all the tasks of the current permutation\footnote{The travelling cost changes among permutations, as the distance covered by robot $r1$ travelling between two tasks $t1,t2$, is different if following the order $\langle initialPosition(r)$, $location(t1)$,$location(t2)\rangle$ or $\langle initialPosition(r)$, $location(t2)$,$location(t1)\rangle$}. 

Third, a nondeterministic action is available at any stage as the robot progresses with its tasks. This action allows the robot to stop for \textit{one time unit}. This \textit{idle} option allows robots to synchronise in the case they need to perform \textit{join} tasks, or wait until previous tasks are completed in case of $ordered$ tasks~(see Figure \ref{fig:compositionMDP}a.3).

\medskip
\noindent
\textbf{MDP model.} Figure \ref{fig:compositionMDP}b shows the composition of an MDP model in the PRISM language. In the first line, \textbf{mdp} declares the type of model. Second, all necessary \textbf{constants} $const$ that are used within the robot modules are declared\footnote{For the description of all constants $const$, we refer the reader directly to Figure \ref{fig:compositionMDP}b.}. Third, the boolean formula \textbf{done} is true when all robots have completed their tasks and the formula \textit{success} is true when no robot fails with their tasks. Fourth, the robot modules $ri$ contains the variables: a)~$r_iorder$, from 0 to \textit{last}$(r_iorder)$; b)~$r_itime$, from 0 to the maximum time available ($TT$); and C)~$r_ifail$ as a Boolean (changing to true when the robot fails a task). For each transition, the guard checks: the order of transitions, that the time does not exceed the available time and that the robot has not failed. When a transition is taken, with a probability of succeeding with the task, the time and order is updated, meaning that the task has been completed; and with a smaller probability, the robot fails and $r_ifail'=true$. Continuing, there is a transition for recovery after failure, where $r_ifail$ becomes false again; and a non-deterministic action $r_iidle$~(feasible if the robot hasn't finished, if it has not failed and if time allows) that adds +1 to the robot timer. Finally, a \textit{travel} reward structure computes the cumulative travelling cost among robots; and an \textit{idle} reward the cumulative time that the robots spend in stand-by~(idling).

For an MDP, we use probabilistic model checking to assess if the selected permutation of tasks is feasible\footnote{The feasibility of a permutation of tasks does not necessarily mean that the robot schedule is feasible. For example, assuming that the permutation of tasks assigned to robot r1 is correct, if the available time is set to 5 and the robot requires 6 time units travelling, the schedule is unfeasible.}, meaning that the non-determinism can be solved to create a \textit{plan} for which each robot knows exactly what to do at every time step and that the plan fulfils all constraints. We check for \textit{feasibility} by assessing:
\begin{equation}
    P_{max=?}[F\; done]
\end{equation}

If there is a path that reaches label \textit{done}, then this formula evaluates to 1, and 0 otherwise. For feasible solutions, we can compute the optimisation objectives depicted in Figure~\ref{fig:compositionMDP}c: finding the minimal probability of success, minimising the idle time and the travelling cost. Note that the first formula evaluates the probability of completing a mission without a failure, which is equivalent to the multiplication of the probabilities of each robot completing its tasks; and the last formula evaluates the travelling cost, which can be computed in advance as describe in Figure~\ref{fig:compositionMDP}a.2.

\subsubsection{Modelling task constraints}
\label{sec:model augmented with task constraints}

\begin{figure}[t!]
    \centering
    \vspace{-5pt}
    \includegraphics[width=0.88\linewidth]{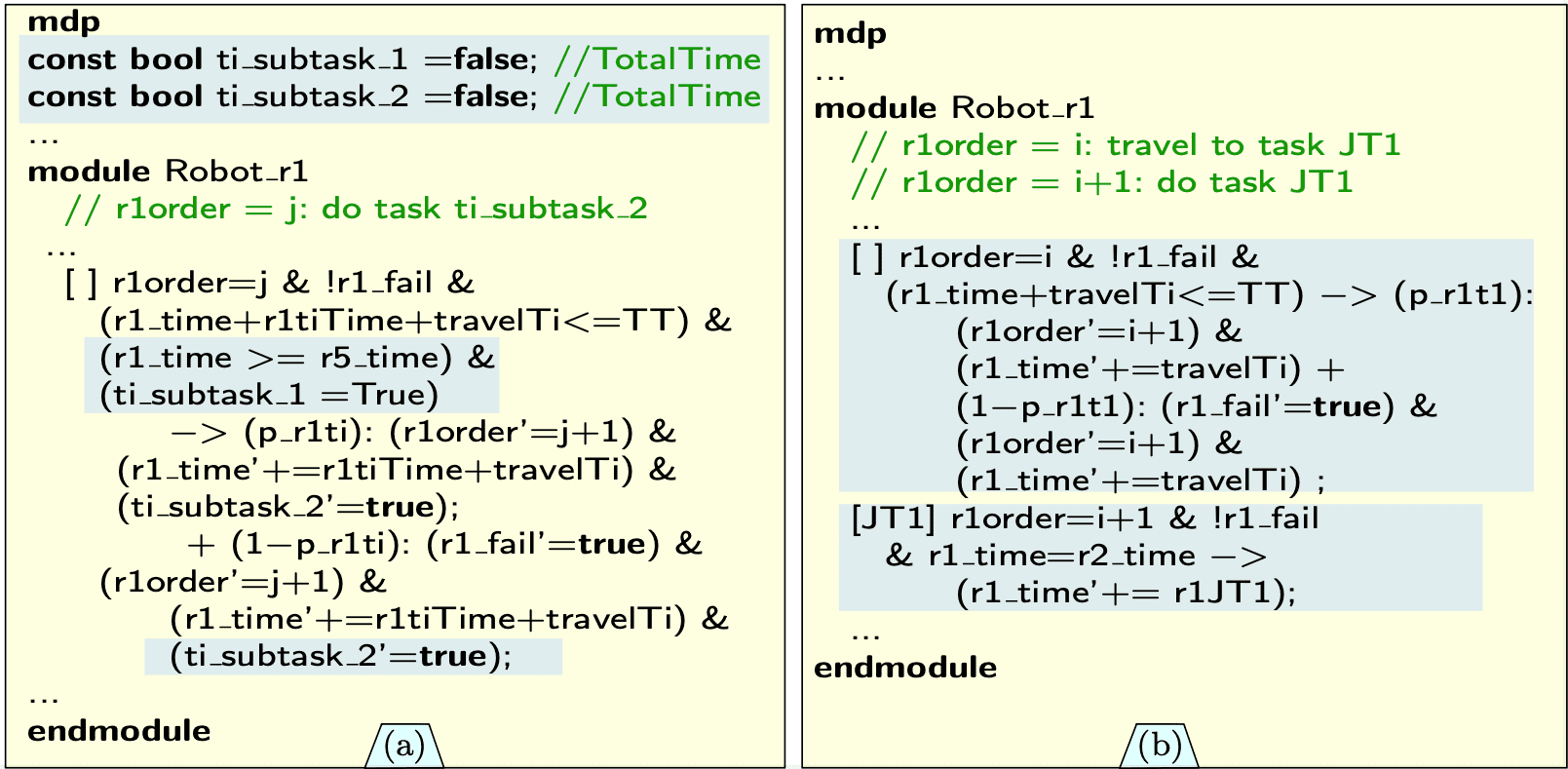}
    \vspace{-14pt}
    \caption{(a) Joint task JT1 in robot r1 written in the PRISM language as two transitions: first travel, then synchronise with other robots via the action label $[JT1]$. All transitions in other modules, with the same action label, ``transition'' at the same time.
    (b) Robot r1 is assigned with the second subtask of the ordered task $ti$. In its $jth$ transition, r1 checks if robot r5 has already completed the first subtask in the past, by checking if r1 time is greater than or equal to r5 time, and $ti\_subtask\_1$ is true. Every constraint subtask is declared as a Boolean constant.
    Changes compared to unconstrained tasks are highlighted in blue.}
    \vspace{-3.5mm}
    \label{fig:joinorderedTask}
\end{figure}

\textbf{Joint tasks}. \textit{Joint tasks} require two or more robots to meet at the same space and time. We model joint tasks using two PRISM transitions. In each module, the robots travel to the location of the joint task, JTi. If the robot is to synchronize in time with the other robots (r$_1$time=r$_2$time=...), synchronous transitions among their modules are possible via a common \textit{action} so that all robots perform the task. Figure~\ref{fig:joinorderedTask}a shows robot $r1$ with the action label $JT1$.

\medskip\noindent
\textbf{Ordered tasks}. 
We model \textit{ordered tasks} in PRISM as single transitions, in the same way as unconstrained tasks. Then, we modify the guard of the $ith$ subtask of the ordered task~($\forall i>1$), to check if its time is greater than or equal to the robot's time assigned with the subtask $i-1$; as well as checking if the subtask $i-1$ itself has been completed. This solution is illustrate in Figure~\ref{fig:joinorderedTask}b.

\subsection{Optimisation of feasible solutions}
\label{sec:approach-selection}

In the previous section, we described the assembly of an MDP model for a group of robots, each of which with a pre-defined sequence of allocated tasks, we used probabilistic model checking to obtain policies satisfying three optimisation objectives: maximisation of the probability of success, and minimisation of the idle time and travelling cost, and we described the assessment of the feasibility of a schedule. We solve the problem of obtaining the Pareto-optimal solutions of this constrained multi-objective optimization problem using a elitist genetic algorithm. Each a chromosome represents an allocation and task permutation as a single gene. The Pareto front of a such a problem comprises all \emph{non-dominated solution}s, i.e., all solutions for which there is no other feasible solution that can improve one of the optimisation objectives without worsening one or more of the others~\cite{durillo2011jmetal}.

Genetic algorithms~(GA) provide an effective approach for solving complex optimisation problems. They operate by encoding a number of solutions (population) whose elements (chromosomes) consist of one or more values, called \textit{genes}, that are genetically bred through the application of the Darwinian principles of survival and reproduction of the fittest solutions, and their recombination (croosover) to create new populations (offspring) that \textit{evolve} towards an optimal solution~\cite{koza1994genetic}. For \acronym, we adopt the widely used NSGA-II (nondominated sorting genetic algorithm II) with chromosomes that encode both the selected allocation and the selected permutation of robot tasks. Furthermore, we use the EvoChecker~\cite{gerasimou2015search,DBLP:journals/ase/GerasimouCT18} framework (which integrates NSGA-II with the PRISM probabilistic model checker)  to obtain the Pareto front of task scheduling solutions.

 \section{Implementation}
\label{sec:implementation}

\begin{wrapfigure}[12]{r}{0.4\textwidth}
    \centering
    \vspace{-50pt}
\includegraphics[width=\linewidth]{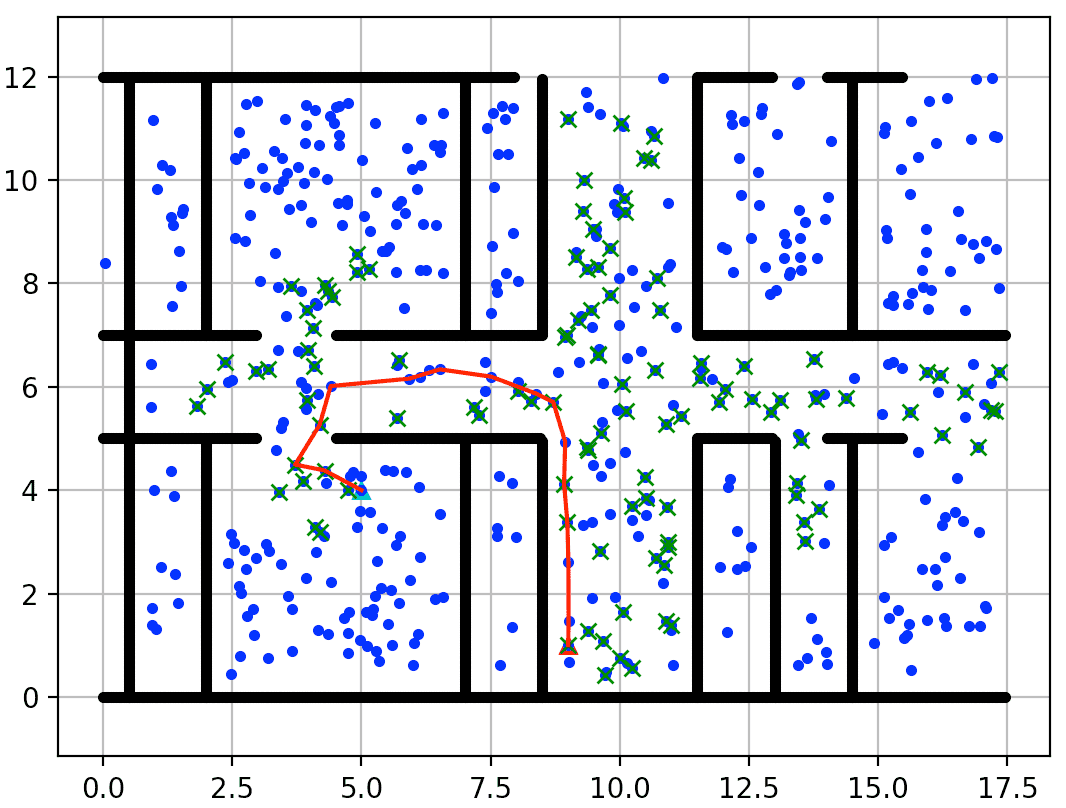}
\vspace{-12pt}
    \caption{Path from the initial location of robot r2 to room 6 found using PRM}
    \label{fig:prm}
\end{wrapfigure}

We developed a prototype tool that implements \acronym. The input is a .mydsl file with the \textit{problem specification} written in our DSL. For the paths in the world model, we use the established probabilistic roadmaps (PRM) path planner to find the distances between locations in the world model, rounding the distances up to the closest integer~(see Figure~\ref{fig:prm})~\cite{kavraki1996probabilistic}.

We used Ecore and Xtend~\cite{bettini2016implementing} to check the DSL syntax and to automatically generate the \textit{Alloy model}. As described in Section \ref{sec:alloy}, Alloy is a declarative language for the specification of systems under a set of constraints supported by the Alloy Analyzer solver tool. Hence, Alloy Analyzer generates up to $N$ allocation files. For each allocation, the \textit{transitive closure} algorithm finds clusters of robots sharing task constraints.

\noindent
The JMetal~5~\cite{nebro2015redesigning} framework was used to solve the optimisation of the permutation of tasks using the NSGA-II algorithm. We use the PRISM model checker~\cite{kwiatkowska2011prism} in the NSGA-II \textit{evaluation} stage, to compute the values of the three objective variables: probability of success, idle time and travelling cost; as well as assessing if a schedule is feasible under the time and space constraints.
The \acronym\ open-source code is available from our GitHub repository at \url{https://github.com/Gricel-lee/Kanoa}.

\section{Evaluation}
\label{sec:evaluation}
In this section, we evaluate \acronym\ applicability using the hospital motivating scenario, and the efficiency of the approach varying the number of tasks, robots and complexity of the tasks.

\begin{wrapfigure}[19]{l}{0.6\textwidth}
    \centering
    \vspace{-16pt}
\includegraphics[width=\linewidth]{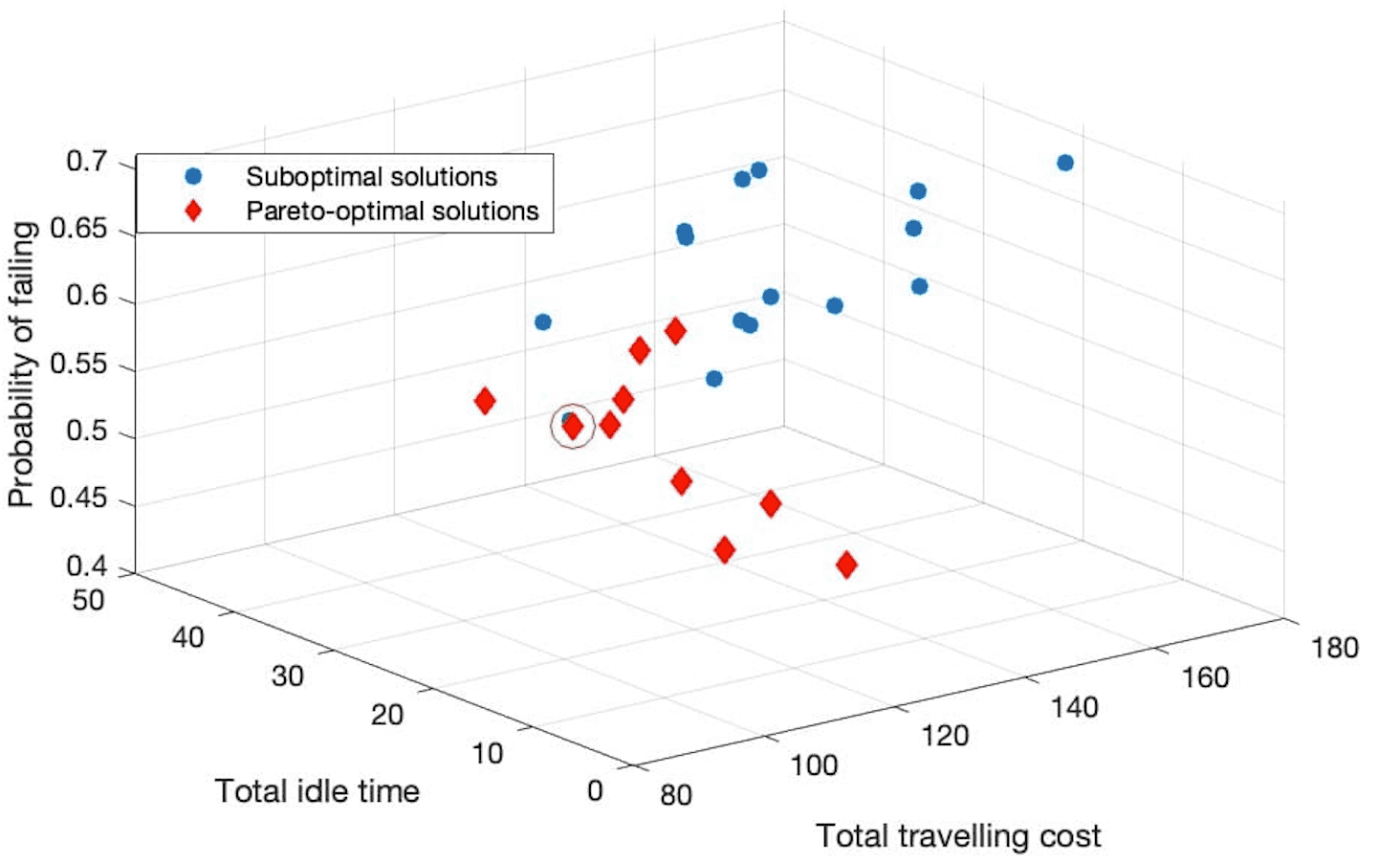}
\vspace{-0pt}
    \caption{Solutions found by \acronym\ for the hospital case study. Pareto-front solutions are coloured in red. The schedule of the solution circled in red, is depicted in Figure~\ref{fig:gantt}, with values corresponding to the first row of Table~\ref{table:solutions}. }
    \label{fig:paretoHosp}
\end{wrapfigure}

\textbf{Application to the motivating example.} To evaluate \acronym, we applied it to the hospital maintenance mission from our motivating example. The DSL-encoded specification of this mission is provided in Figure~\ref{fig:toolDSL} earlier in the paper. This specification was supplied as input to \acronym\ tool. Internally, the Alloy analyzer is configured to generate 30 allocations models. One of these models is depicted in Figure~\ref{fig:workspace} earlier in the paper.

\begin{figure}[t]
    \centering
    \includegraphics[width=0.8\linewidth]{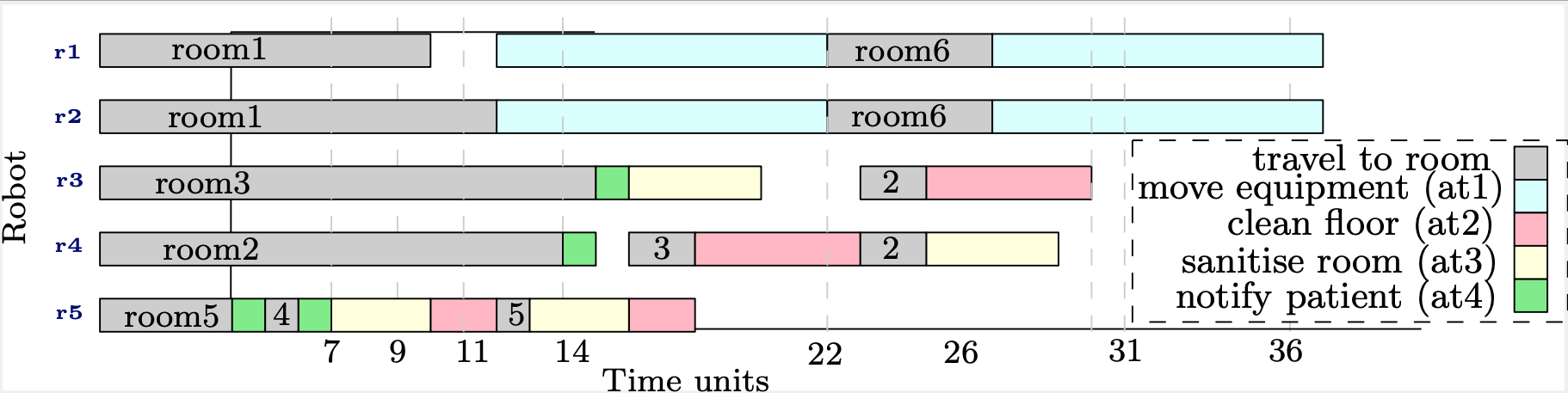}
    \vspace*{-3mm}
    \caption{Example of a scheduled plan synthesized by \acronym. All results were obtained on a MacBook Air computer with Apple M1 processor, macOS Monterey 12.3.1, and 8~GB of memory.}
    \label{fig:gantt}
\end{figure}

Pre-scheduling was then applied to each of these task allocations, yielding between two and three subsets of robots with one to three robots in each. The \acronym\ genetic algorithm was configured to run for 5 iterations with a population of 50 chromosomes. Each of the 30 allocation was allowed to compute 20 permutations, giving a total of 600 possible chromosomes. The idle time, travelling cost and probability of failing to execute the mission for the Pareto-optimal solutions evaluated across the task allocations are shown in Figure~\ref{fig:paretoHosp}, and one of these solutions is detailed along the timeline in Figure~\ref{fig:gantt}. We note that: (i)~the failure probability was used instead of the success probability because multiobjective optimisation genetic algorithms operate by minimising all objectives; and (ii)~that the values obtained for the failure probability are unacceptably high, which is due to using randomly selected failure probabilities for the individual robots in our experiments. Nevertheless, neither of these issues affects the validity of the experimental results, which were only intended to assess the ability of \acronym\ to produce Pareto-optimal plans for MRS missions.

From the Pareto-optimal solutions detailed in Table~\ref{table:solutions}, four solutions (rows 1, 3, 6 and 10) achieve very low travelling cost and the idling time at the expense of high failure probabilities. The rest of the Pareto-optimal solutions correspond to solutions that achieve much lower probabilities of failure (with values below 0.55) in exchange for completing the mission with additional travelling cost and idling time.

\vspace{-1pt}
\begin{wraptable}{l}{0.5\textwidth}
\vspace{-12pt}
\centering
\caption{Pareto-optimal solutions}
\label{table:solutions}

\footnotesize
\sffamily
\begin{tabular}{l|l|c|c|c|c|}
\cline{2-6}
 & \textbf{\begin{tabular}[c]{@{}c@{}}Alloc\\ation\end{tabular}} & \textbf{\begin{tabular}[c]{@{}c@{}}Permu\\tation\end{tabular}} & \textbf{\begin{tabular}[c]{@{}c@{}}Probability\\ of failure\end{tabular}} & \textbf{\begin{tabular}[c]{@{}c@{}}Idling\end{tabular}} & \textbf{\begin{tabular}[c]{@{}c@{}}Travel\end{tabular}} \\ \hline
\multicolumn{1}{|l|}{1} & 17 & 9 & 0.63423811 & 6 & 80 \\ \hline
\multicolumn{1}{|l|}{2} & 5 & 5 & 0.51937429 & 39 & 117 \\ \hline
\multicolumn{1}{|l|}{3} & 17 & 5 & 0.63423811 & 5 & 84 \\ \hline
\multicolumn{1}{|l|}{4} & 16 & 5 & 0.53284079 & 31 & 126 \\ \hline
\multicolumn{1}{|l|}{5} & 11 & 1 & 0.41844289 & 28 & 137 \\ \hline
\multicolumn{1}{|l|}{6} & 21 & 1 & 0.69771745 & 2 & 84 \\ \hline
\multicolumn{1}{|l|}{7} & 11 & 7 & 0.41844289 & 21 & 145 \\ \hline
\multicolumn{1}{|l|}{8} & 9 & 8 & 0.47131171 & 29 & 132 \\ \hline
\multicolumn{1}{|l|}{9} & 9 & 1 & 0.47131171 & 22 & 135 \\ \hline
\multicolumn{1}{|l|}{10} & 12 & 8 & 0.70993088 & 1 & 88 \\ \hline
\end{tabular}
\end{wraptable}

Table~\ref{table:solutions} shows the number of allocations and permutations that the Pareto solutions belong to. It is interesting that there is not a single allocation of tasks to robots that seem to perform better than the rest. For instance, allocations 9, 11 and 17 each occur twice. Also notice that the probability of success is the same among the same allocations, however, the idling time and travelling costs change as the permutation changes.

\noindent
Figure~\ref{fig:gantt} shows a schedule from the Pareto front; in this schedule, all robots were deployed. Robots r1 and r2 first visit room1, perform task at1 (moving medical equipment), travel to room2, and again perform at1. In this case, \acronym\ found that deploying robots r3 and r4 to clean rooms 2 and 3, and robot r5 to clean rooms 4 and 5 is a Pareto-optimal schedule. Notice that these three cleaner robots travel between the rooms back and forth~(see also Figure~\ref{fig:workspace}). This is because the distance between rooms 2 to 3 and 4 to 5, is relatively small that it represents a better solution that travelling between other rooms. Although a better solution, where robots avoid repeating rooms, was not found (given the number of permutations and allocations, population and evaluation size), the synthesized solution complies with all requirements and, besides GA, no heuristic guidance was used in the searching process.

\begin{figure}[t]
\centering
    \includegraphics[width=0.8\linewidth]{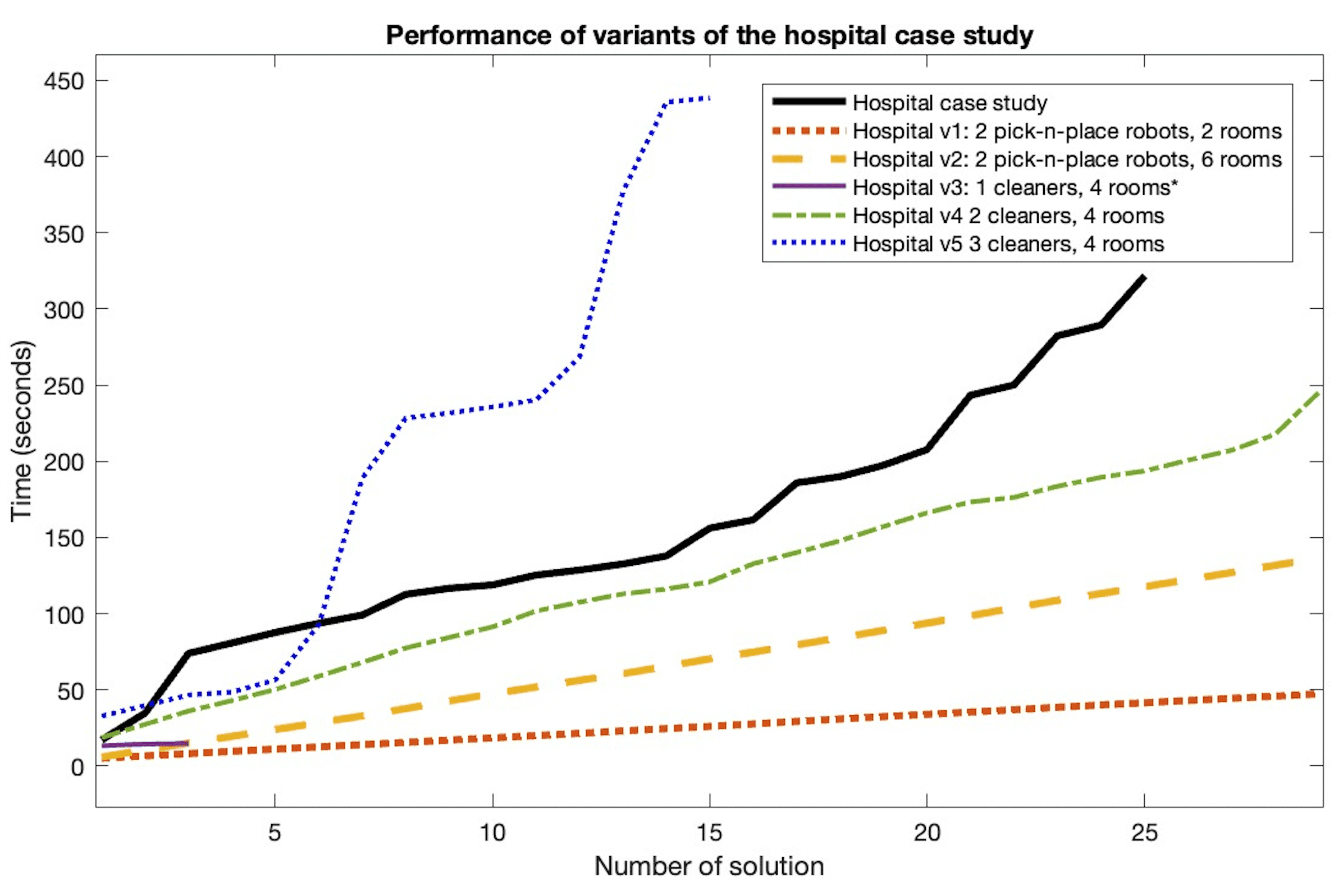}
    \caption{Scheduled plan synthesized by \acronym; all solutions were computed with 30 allocations, 20 permutation of tasks, with population size 50 and 5 evaluations (except variants v3, which was tested with 500 permutations)}
    \label{fig:performance}
\end{figure}

\medskip
\noindent
\textbf{Efficiency.} We also carried out preliminary experiments to assess the execution time of \acronym. We tested five variants of the hospital scenario in order to assess the impact that (a)~adding new tasks and (b)~adding more robots have on the time required to find and assess a solution; and (c)~the impact that the complexity of the task constraints have on the overall performance. The results obtained for all five variants and for the original case study are shown in Figure~\ref{fig:performance}. The experiments were run until 30 feasible solutions were found or until the time surpassed 300 seconds, whichever occurred first.

Variants v1 and v2 involved using the two pick-and-place robots to move surgical equipment in multiple rooms. For variant v1, they must visit two rooms, and for variant v2 there were six rooms to service. The time to get one feasible solution was 9ms for both variants. However, the time for variant v3 increases to a mean of 4.7s per solution, while for variant v4 \acronym\ finds a solution every 1.33s. This is an increase of approximately 3.53 times the time to assess a solution when the number of tasks tripled. As v1 and v2 plots show a linear increase, this means that we expect the time to increase approximately 1.17 times for each task added to this hospital scenario.

The third, fourth and fifth variants (v3, v4, and v5) correspond to one, two and three robots available, respectively, with the mission to clean patient rooms 2 to 5. For variant v3, we only found 3 solutions as only one robot has all tasks assigned, making it difficult to synthesize plans that comply with the time limit. The time to find the solutions is 14.86 seconds, this is understandable as, for most permutations, PRISM is called only to check for feasibility and discarded as they do not comply with the requirements. Variants v4 and v5, increasing v3 by one and two robots, show a considerable increase in the time to find a solution. This is understandable as the complexity of the PRISM models increases, taking more time for the PMC to check for feasibility and to return the probability of failure, idling time and travelling cost.

\section{Related Work}
\label{sec:related}

Work in other domains, such as manufacturing~\cite{wong2006dynamic} and distributed computer systems~\cite{wolf2008soda} has influenced the algorithms used to solve allocation and scheduling problems in robotics. Some of the most broadly researched categories are mixed-integer linear programming (MILP), and hybrid methods such as MILP constraint programming and MILP depth-first search heuristics~\cite{chen2009project}. These approaches exhibit desirable traits such as scalability, but often require crafting  application-dependent heuristics by domain-experts to become scalable, simplify task complexity to concurrent and temporal constraints of simple tasks, and do not consider uncertainty~\cite{wang2020learning,gombolay2013fast,calinescu2013emerging}.

On the other hand, model checking has successfully been applied in MRS to provide guarantees for the synthesised robot schedules. To solve the allocation and scheduling problems, these can be modelled together or separately. There is not a clear preference adopted by the research community. For instance, some research only focuses on one of these problems.  Menghi et al.~\cite{menghi2018multi} used model checking for scheduling when only incomplete knowledge is available; and Yu et al.~\cite{yu2021distributed} consider constraints due to the robots sharing a room or hall, and limited local information for re-planning online. When allocation and scheduling are modelled together, we notice the introduction of optimisation metrics. For instance, Antlab~\cite{gavran2017antlab} ranks the schedules by travelling cost, and Ulusoy et al.~\cite{ulusoy2011optimal} optimise the time for the completion of all tasks. In terms of probabilistic model checking, Lacerda et al.~\cite{lacerda2019probabilistic} presents a single-robot planning problem to perform a series of tasks. However, they consider the tasks ordered in advance, and PMC only used to select a path amount pre-define paths. Others use PMC for the synthesis of MRS controllers without considering~\cite{fraser2020collaborative}, inferring the allocation and scheduling as a prerequisite.

\acronym\ extends our preliminary work from~\cite{vazquez2021automated} (where we advocate the use of this type of approach in a doctoral symposium paper) and~\cite{vazquez2021scheduling} (where we assess the feasibility of applying Alloy analyzer and PRISM), and tackles the allocation and scheduling of tasks separately. The information about the robot capabilities and mission tasks is used for both the allocation and the scheduling of tasks, allowing us to modify how the schedules are done by grouping robots from the allocations found, thus reducing the complexity of the models used in the \acronym's scheduling stage. We also assume that the path planning is performed in advance, as adding the planning into the model (e.g., as in \cite{gavran2017antlab}) requires adding each intermediate point, increasing the states and transitions of the model and reducing scalability. In further work, we will reduce the state explosion problem by adding middle points where necessary, for instance, in the hospital halls to avoid two robots entering the same location at the same time.
Other interesting MRS characteristics were also found in the state-of-the-art. Antlab~\cite{gavran2017antlab} captures dynamic collision avoidance, and MALTA~\cite{missionplanning} models road conditions that may slow down robots. MRS must deal with a large range of uncertainty sources.

Although the ultimate goal for \textit{multi-robot planning, scheduling and allocation} may be the creation of a single tool that is applicable to all the variants of MRS applications,
this will be very challenging to achieve (and may not be feasible) because there are too many applications with different critical characteristics to model. For example, for search-and-rescue missions, the robots must be able to work in unknown environments; while in a factory, robots may have a map in advance, although they must be able to avoid staff and cooperate with operators. In further work, we would like to incorporate further task constraints, such as time windows (for example, serving meals to the patients between 13:00 and 14:00), and consecutive tasks (tasks that must be perform one after the other).

\section{Conclusion}
\label{sec:conclusion}

We introduced a new end-to-end approach for the task allocation and scheduling of multi-robot missions comprising joint and ordered tasks that need to be executed by teams of robots with different capabilities. Three different optimisation objectives and a feasibility constraint were use to find Pareto-optimal solutions. We support our end-to-end approach (from the high-level problem definition to the synthesis of robot plans) with an open-source tool and a preliminary evaluation of a hospital case study. We presented five different variants of the MRS mission from this case study, assessing the impact of adding new tasks and robots to the MRS mission. 

In future work, we will explore options for improving the scalability of \acronym, and for gracefully degrading the constraints defined in the \acronym\ problem specifications when no plan can be found when the initial constraints are taken into account. 
We also aim to enhance \acronym\ with self-adaptation capabilities, enabling the dynamic evolution of the generated plans after disturbances such as the failure of a robot, the failure of a robot's sensors/actuators, or a delay in the execution of a task by one of the robots. Moreover, longer runs will be used in the evaluation of our future work to establish whether a longer search produces significantly better solutions.
Additionally, we will explore the inclusion of spatial restrictions, such as ``only one robot can pass through a hall at a time'' or ``no more than two robots are allowed in a room at the same time''. Moreover, as we deal with autonomous mobile robots, we need to introduce an event when the battery is low so that the robot can reach a charging station.

Understanding the computational complexity of our approach is another area of future work for the project. Analysing the complexity of the approach is nontrivial, as it depends on the Alloy Analyser solver, PRISM engine and JMetal configuration, and, as described in Section~\ref{sec:evaluation}, it also depends on the number of robots, number of tasks and task dependencies.

\bibliographystyle{eptcs}
\bibliography{generic}

\end{document}